\documentclass[fleqn,10pt]{wlscirep}
\usepackage[utf8]{inputenc}
\usepackage[T1]{fontenc}

\usepackage{graphicx}
\usepackage{subcaption}
\usepackage[labelfont=bf]{caption}

\usepackage{nameref}

\title{Enhancing Traffic Safety Analysis with Digital Twin Technology: Integrating Vehicle Dynamics and Environmental Factors into Microscopic Traffic Simulation \thanks{This manuscript has been authored in part by UT-Battelle, LLC, under contract DE-AC05-00OR22725 with the US Department of Energy (DOE). The publisher acknowledges the US government license to provide public access under the DOE Public Access Plan {(http://energy.gov/downloads/doe-public-access-plan)}.}}

\author[1,+,**]{Guanhao Xu}
\author[1,+]{Jianfei Chen}
\author[1,+]{Zejiang Wang}
\author[1,+]{Anye Zhou}
\author[2,+]{Max Schrader}
\author[2,+]{Joshua Bittle}
\author[3,+]{Yunli Shao}
\affil[1]{Buildings and Transportation Science Division, Oak Ridge National Laboratory, Oak Ridge, TN 37831, USA}
\affil[2]{Department of Mechanical Engineering, The University of Alabama, Tuscaloosa, AL 35487, USA}
\affil[3]{School of Environmental, Civil, Agricultural, and Mechanical Engineering, University of Georgia, GA 30602, USA}
\affil[**]{xug1@ornl.gov}

\affil[+]{these authors contributed equally to this work}


\begin{abstract}
Traffic safety is a critical concern in transportation engineering and urban planning. Traditional traffic safety analysis requires trained observers to collect data in the field, which is time-consuming, labor-intensive, and sometimes inaccurate. In recent years, microscopic traffic simulation, which simulates individual vehicles' movements within a transportation network, have been utilized to study traffic safety. However, microscopic traffic simulation only focuses on traffic-related factors, such as traffic volume, traffic signals, and lane configurations, neglecting vehicle dynamics and environment-related factors like weather and lighting conditions, which can significantly impact traffic safety. In light of this, this paper explores the application of digital twin technology in traffic safety analysis, integrating vehicle simulators, which consider vehicle dynamics and environmental factors, and microscopic traffic simulators, which simulate the operations of traffic flow, for enhanced safety evaluations. Various scenarios, including different weather conditions and visibility levels, are simulated using a digital twin of a road segment in Tuscaloosa, Alabama. The simulations employ Surrogate Safety Measures (SSMs) like Time to Collision (TTC) and Deceleration Rate to Avoid a Crash (DRAC) to assess safety under varying conditions. The results demonstrate that traffic digital twin can identify potential safety issues that traditional microscopic simulation cannot, providing insights for improving traffic control strategies and  transportation infrastructure to enhance traffic safety.

\end{abstract}
\begin{document}

\flushbottom
\maketitle

\thispagestyle{empty}

\section*{Introduction}

\subsection*{Traffic safety measurement} 
Traffic safety is a critical concern in transportation engineering and urban planning and is often measured by crash frequency and severity. Crash frequency refers to the number of crashes occurring within a specific time period or location, while crash severity categorizes the crashes based on the level of damage or injury sustained, such as property damage only, minor injuries, major injuries, or fatalities \cite{wang2018combined}. The modeling of crash frequency often relies on statistical models, such as Poisson, Negative Binomial, or Zero-Inflated models, to account for the discrete nature of crash data and over-dispersion issues\cite{dong2014multivariate,mannering2014analytic,lord2010statistical,abdel2000modeling,poch1996negative}. Likewise, the crash severity is often analyzed using regression models, such as ordered probit or logit models, to understand the factors influencing the severity levels of accidents \cite{kockelman2002driver,savolainen2011statistical,mannering2014analytic,rahim2021deep}. Recent advancements have seen the integration of machine learning techniques to analyze crash frequency\cite{zhang2020ensemble,zhou2023integration,wen2022interpretability,theofilatos2019comparing} and or crash severity \cite{wen2021applications,iranitalab2017comparison,almamlook2019comparison}. However, the availability and completeness of historical crash data are often a challenge when developing crash-based statistical models.  For instance, crash data are typically unavailable for new roads or existing roads with a safety treatment applied on. Moreover, the quality of crash data is often compromised due to underreporting, misclassification, or inconsistent data collection practices\cite{imprialou2019crash,abdulhafedh2017road}. These issues hinder the application of traditional crash-based statistical models, which heavily rely on historical crash information to identify trends and evaluate the impact of safety interventions.

In response to these challenges, the development of Surrogate Safety Measures (SSMs) has gained prominence as an alternative approach for traffic safety evaluation\cite{vedagiri2015traffic,tarko2018surrogate,johnsson2018search,wang2021review,sarkar2024review}. SSMs are indirect indicators of safety performance, such as traffic conflict techniques, near-crash events, or other observable behaviors that have a statistical relationship with actual crashes\cite{arun2021systematic}.   These measures can provide valuable insights into the safety performance of roadways, especially in situations where conventional crash data are sparse or unreliable. 
Among all SSMs in the literature, the traffic conflict technique (TCT) is one of the most widely used measures \cite{glennon1977critique,parker1988traffic,chin1997measurement,mahmud2018reviewing,zheng2014traffic,arun2021systematic}. There are two types of definitions for traffic conflict in the literature: evasive actions-based traffic conflict and temporal/spatial proximity-based traffic conflict. Traffic conflicts based on evasive actions are situations where road users have to take evasive actions to avoid a collision\cite{parker1989traffic}.  However, as crashes often occur without prior evasive actions and not all evasive actions indicate a dangerous situation, a good correlation between crashes and conflicts may not exist if conflicts are defined solely based on observed evasive actions\cite{gettman2008surrogate}. By comparison, traffic conflicts based on temporal/spatial proximity are situations where there is a collision risk if the road users’ movements remain unchanged\cite{amundsen1977proceedings}. This definition allows interpretable and easily understood quantitative measurements in terms of closeness to the collision\cite{mahmud2018reviewing}. Some widely used temporal proximity-based traffic conflict SSMs include time to collision, post encroachment time, time headway, etc. Some widely used spatial proximity-based traffic conflict SSMs include proportion of stopping distance, spacing, unsafe density, etc. There are also some proximity-based traffic conflict SSMs related to deceleration, such as deceleration rate to avoid a crash and brake rate\cite{gettman2003surrogate}.

To obtain these SSMs, traditional traffic conflict studies require trained observers to collect conflict data in the field, which is time-consuming, labor-intensive, and sometimes inaccurate \cite{wang2018combined}. As a result, traffic conflict techniques in conjugate with microscopic traffic simulation have been widely adopted in recent years\cite{yang2010application,wang2014evaluation,astarita2019surrogate,astarita2020surrogate}. Microscopic traffic simulation, which simulates individual vehicles' movements within a transportation network, offers the advantage of controlling various traffic parameters and conditions\cite{xu2023non}, enabling a comprehensive exploration of potential safety issues and the effectiveness of various interventions\cite{archer2005indicators}. However, microscopic traffic simulation focuses on traffic-related factors, such as traffic volume, traffic signals, and lane configurations, neglecting environment-related factors like weather and lighting conditions, which can significantly impact traffic safety. In addition, microscopic traffic simulations use car-following models to simulate vehicle movements. While these models are adequate for simulating interactions between vehicles, they simplify the complex dynamics of vehicles (to reduce computational complexity), such as acceleration and braking, which are closely related to traffic safety, especially in scenarios involving rapid acceleration, deceleration, or sharp turns. Recently, emerging technologies such as digital twin have been developed and can potentially be used for traffic safety research to address these limitations.

\subsection*{Digital twin} 
A digital twin is a digital replica of a related physical object in the real world, connecting the two parts through real-time two-way communication\cite{wright2020tell}. On the one hand, data from physical objects is used to update the digital twin. On the other hand, control commands from the digital twin can be transmitted to the physical object to change its state\cite{irfan2022towards}. Real-time synchronization between the digital and physical worlds enables digital twins to represent objects that change over time accurately. Moreover, it enables the digital twin to predict the state of the corresponding physical object under the influence of digital side commands and evaluate the effectiveness of different control strategies.

Digital twins can play an important role in improving the safety, efficiency, mobility, and sustainability of modern transportation and vehicle systems. In terms of transportation systems, created a continuously synchronized digital twin was created for the Geneva Motorway\cite{kuvsic2023digital}. The aim was to predict traffic conditions before safety-critical traffic management systems are deployed in the real world. A pipeline for constructing a digital twin was proposed and used to build digital twin of Barcelona's urban traffic during peak hours\cite{argota2022getting}. For connected and autonomous vehicles (CAVs), digital twins play a critical role in controller design, validation, and testing. For example, use cases for digital twins were summarized\cite{schwarz2022role}, focusing on CAV testing and performance evaluation. In addition, CAV-related digital twin applications were further classified into four categories\cite{bhatti2021towards}:
(A). Predictive mobility and autonomous motion control, where CAV digital twins provide training and test data sets for various navigation, planning and control algorithms.
(B). Advanced driver assistance systems (ADAS), where ADAS digital twins mix historical and real-time data for driver behavior prediction and system control.(C). Vehicle health monitoring and management, in which digital twins of various vehicle components reflect the health status and probabilistic failures of vehicle systems and perform condition-based maintenance to prevent the degradation of safety-critical systems.
(D). Battery management system (BMS) and smart charging, where BMS digital twins are used for battery and charging system diagnosis and prediction.

\subsection*{Barriers and Contributions}

Today's traffic system is a system-of-systems that includes (1) road geometry and environment, (2) road users such as vehicles, cyclists, pedestrians, etc., and (3) traffic controllers, such as roadside units, traffic lights, etc. It is challenging to fully observe real-world traffic systems to measure states of all these elements above. Digital twin offers a promising solution by creating a digital replica of the physical traffic environment, thereby informing traffic managers of current traffic conditions and enabling timely traffic management decisions to improve system level performance. For example, digital twins can continuously monitor infrastructure health, identifying potential issues such as road damage or traffic signal failures before they cause accidents\cite{ye2019digital}. As another example, digital twins can incorporate the behavioral patterns of pedestrians and cyclists, enabling safer infrastructure design for optimal visibility and safety\cite{wang2023towards}. The success of applying digital twins to transportation systems hinges on developing high-quality simulation environments that replicate diverse roads and driving conditions and the complex behaviors of road users and traffic controllers in the real-world. Existing traffic digital twins are mostly developed based on microscopic traffic simulation tools\cite{keler2023calibration,saroj2022optimizing}, which can simulate traffic-level performance with high fidelity. However, these tools lack the ability to simulate factors that significantly impact traffic safety, including environment factors (weather and lighting conditions) and detailed vehicle and powertrain dynamics. 

In this paper, we present a traffic digital twin framework for traffic safety analysis that includes high-quality simulation of vehicle dynamics and environmental factors. This is achieved through a co-simulation architecture combining detailed vehicle and driving environment simulators with microscopic traffic simulators. Compared to current traffic digital twins that use only microscopic traffic simulations, our innovative digital twin approach addresses the need for considering detailed vehicle dynamics and environment factors, providing a high-fidelity and comprehensive understanding of traffic safety. The resulting traffic digital twin helps identify potential road safety concerns under various conditions and allows for the examination of complex interactions between vehicles, transportation infrastructures, and environmental impacts. This enhances the ability of transportation agencies to evaluate and improve traffic safety. For instance, using the developed digital twin for a real-world road in this study, we identified a potential crash at a signalized intersection when road friction decreases due to weather conditions such as rain or snow. This information can be used to provide real-time feedback to traffic signal controllers, recommending a longer green time to let the driver traverse the intersection or a longer yellow time to allow drivers more time to react and stop safely at the intersection.


\section*{Methods}

\subsection*{Workflow of traffic digital twin generation}

\begin{figure}
    \centering
    \includegraphics[width=1\linewidth]{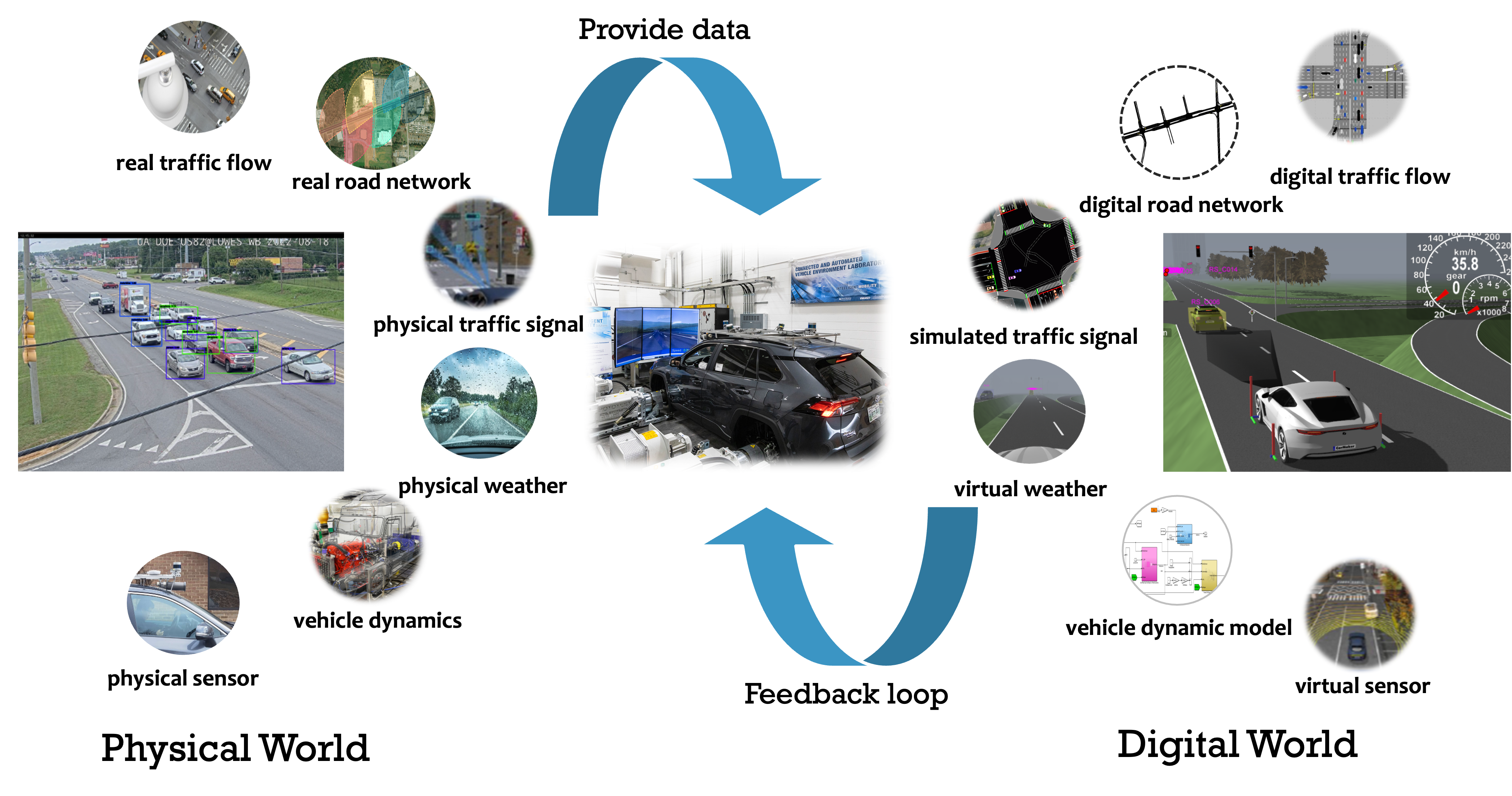}
    \caption{Digital twin concept for traffic and vehicle co-simulation}
    \label{fig:enter-label}
\end{figure}

To analyze traffic safety with high realism, we have developed a digital twin framework for the physical traffic scenario, as shown in Figure \ref{fig:enter-label}. This framework seamlessly integrates road geometry, environment, road users, and traffic controllers in a synchronized manner with both vehicle and driving environment simulators and microscopic traffic simulators.  In this work, we use the \textit{IPG CarMaker} (refer to as CM later in the paper)\cite{carmaker2009} as the vehicle and driving environment simulator and \textit{Simulation of Urban Mobility} (SUMO)\cite{sumo} as the microscopic traffic simulator. CM, a commercial vehicle simulator by IPG automotive, provides high-fidelity vehicle models including tires, powertrain, chassis dynamics, and sensors. It also contains a 3D environment that can reproduce road geometry, surface conditions, and surrounding environment and lighting conditions. SUMO is an open-source, highly portable, microscopic, and continuous traffic simulation software designed to simulate large transportation networks. It supports the simulation of different road users, including pedestrians, and can replicate real-world traffic-related conditions such as traffic volume, density, and traffic signals\cite{sumo}.

For the physical world of interest, the digital twin replicates the actual roadways and all road users. In the co-simulation framework, all road users are simulated in SUMO, while CM simulates selected ego vehicles with detailed vehicle dynamics and vehicle-environment interactions. For example, ego vehicles are connected and automated vehicles that share real-time observations with the digital twin and receive traffic management recommendations from it. Although the number of simulated vehicles differ in SUMO and CM, both share identical road geometry and traffic controller of the physical world. The road geometry is generated in OpenDRIVE format and imported into both simulators. The traffic controllers are defined in SUMO and exported to CM environment. The co-simulation of SUMO and CM requires real-time synchronization of ego vehicles from CM with all other road users in SUMO. The synchronization architecture is designed using a sequential workflow. By default, the simulation time step for SUMO is 0.1 seconds, while for CM it is 0.001 seconds. Every 0.1 seconds, the states of ego vehicles are sent to SUMO using customized TCP/IP messages. Meanwhile, SUMO runs one step forward and returns updated states of road users to CM. For a detailed discussion of the synchronization architecture, refer to \cite{shao2022real, shao2023evaluating}. To alleviate the computational burden of sharing information between the SUMO and CM in real-time, we only send data for vehicles in the vicinity of ego vehicles from SUMO to CM. The synchronization interface has been published as open-source software named the FIXS interface \cite{shao2023real}, a tool by Oak Ridge National Lab (ORNL) that can be used not only for traffic digital twins as shown in this work, but also other anything-in-the-loop (XIL) co-simulation with microscopic traffic simulator and vehicle and driving environment simulator.

Modern signalized intersections are equipped with connected infrastructures that measure real-world traffic conditions, such as traffic volumes, turning rates, routes, speed, etc. These data are often stored in a database at traffic management centers (TMC). The proposed traffic digital twin will access these databases to use the real-world traffic information to update the traffic conditions. An example of the data collection and digital twin creation is discussed in the next section. The digital twin will analyze traffic safety under the real-world traffic conditions, and the results can be sent back to the TMC to improve traffic management. Example applications of the digital twin are discussed in the Results section. 

\subsection*{Case study location} 

A traffic digital twin is built for a segment of a divided state highway in Tuscaloosa, Alabama. This arterial route serves as a crucial east-west traffic corridor, connecting residential areas in the west to the urban center of Tuscaloosa in the east. The highway experiences significant traffic flow, especially during peak commuting hours. The study area includes three signalized intersections, which are modeled in the simulation using SUMO's NEMA controllers~\cite{nema_controller}. To accurately represent the traffic volume in the simulation, data from the Econolite EVO system's zone-based radar detection is utilized, which is stored in a central SQL server using an enumerated format~\cite{Indiana_Traffic_Signal_logger}.

The traffic flows for the network are derived from the location-resolved volume counts stored in the SQL database. This process is carried out using the tool \textit{routeSampler}, which employs integer linear programming to identify a set of vehicle trips through the network that satisfy the observed volume counts. These traffic data from the SQL database will be shared to the digital twin to replicate the real-world traffic conditions. The representative day for trip generation is selected as July 24th, 2023, based on the guidelines outlined in the 2019 Update to the 2004 USDOT Traffic Analysis Toolbox Volume III: Guidelines for Applying Traffic Microsimulation Modeling Software~\cite{usdot_calibration}. This date is identified from the study period spanning from July 1st to August 31st, 2023. The simulation focuses on the time range from 5:30 AM to 12:00 PM on the target day, encompassing the morning rush hour and periods of mid-day traffic. To ensure the reliability of the simulation, the simulated volume counts are validated against the volume-calibration based criteria provided in the USDOT guidelines, confirming that they fall within the specified variational envelopes.

\subsection*{Simulation parameters}

In this study, IPG CarMaker 11.0.1 HIL version, and SUMO 1.18.0 are used. A pre-defined electrical vehicle (Demo IPG CompanyCar EV 2) from CM is used as the ego vehicle, which includes two electrical motors and tire model of 255-35r20. The ego vehicle's surrounding traffic are set to the same model for ease of measurement. 

Additional virtual object sensors are added to the vehicle to sense the front vehicle's existence, speed and distance. The detection distance can be configured as needed, for friction related simulation, the default range is set as 150 meters. For visibility related simulations, the detection range is set accordingly to simulate the driver's visibility under different weather conditions. An Intelligent driver model (IDM) is modeled to generate reference speed for the ego vehicle, and a vehicle low level controller is modeled to translate to gas and brake pedal control. 

\subsection*{Simulated scenarios}
To illustrate how traffic digital twins can improve the analysis of traffic safety, we first simulate the digital twin with dry road surfaces and during daytime conditions. This scenario serve as baseline scenario highlighting the importance of considering vehicle dynamics in traffic safety analysis which traditional microscopic traffic simulations fail to.

Second, we simulate scenarios with different friction coefficients corresponding to various weather conditions. The location for which we built the digital twin consists of asphalt roads. Therefore, higher friction coefficients ($\mu = 0.7-0.8$) represent dry conditions while lower friction coefficients ($\mu = 0.4-0.5$) represent rainy conditions \cite{novikov2018study}. Friction coefficients lower than 0.4 can be used to simulate snowy conditions. By varying the friction coefficients to mimic different weather scenarios, we can observe how changes in road surface conditions due to weather affect vehicle behavior and crash risk. This can emphasize the digital twin's ability to model and predict the effects of different weather-induced friction levels and thereby identify critical safety concerns and informing the development of targeted traffic control strategies, such as variable speed limits, to mitigate the risk of crashes.

Finally, we conduct simulations with varying visibility distances ($L_v$) to mimic different environmental conditions such as nighttime driving or foggy weather. Visibility is a key factor in traffic safety, as reduced visibility can significantly influence a driver's ability to perceive and react to other road users (e.g., vehicles in front, cyclists, pedestrians) and transportation infrastructures (e.g., stop sign). By examining these conditions, we aim to show how changes in visibility impact safety illustrating the effectiveness of digital twins in modeling the influence of environmental factors on road safety and providing critical insights for developing traffic control strategies to mitigate risks of crashes.

\subsection*{Measurement of traffic safety} 
In this papers, two Surrogate Safety Measures (SSMs), Time to Collision (TTC) and Deceleration Rate to Avoid a Crash (DRAC), are used to evaluate the traffic safety in the case study.
Time-to-collision (TTC) is a widely used metric that helps understand potential collision risks on the road. TTC represents the time remaining before a collision occurs if both vehicles maintain their current course and speed \cite{vogel2003} and is defined by \eqref{eq:TTC}, where \( S_{t,i} \) is the spacing between vehicle \(i\) and its leader at time \(t\), \(v_{t,i}\) is the speed of vehicle \(i\) at time \(t\),\(v_{t,i+1}\) is the speed of the leader vehicle in front of vehicle \(i\) at time \(t\). A TTC of less than 3 $sec$ is often considered as a traffic conflict, a near-miss situation that, although it doesn't result in a crash, indicates a significant risk of collision. 
\begin{align}
\label{eq:TTC}
TTC_{t,i} = \frac{S_{t,i}}{v_{t,i}-v_{t,i+1}} 
\end{align}

Deceleration Rate to Avoid a Crash (DRAC) is another critical measurement that is widely used in traffic safety analysis. DRAC refers to the rate at which a vehicle must decelerate to avoid a collision with an obstacle or a leader vehicle in its path considering its current speed and the distance to the obstacle or leader vehicle \cite{gettman2003surrogate}. It is given by: \eqref{eq:DRAC}.
\begin{align}
\label{eq:DRAC}
DRAC_{t,i} = \frac{(v_{t,i}-v_{t,i+1})^2}{2S_{t,i}} 
\end{align}

In our case study, we calculate and monitor the TTC and DRAC of of an ego vehicle operating within a constructed digital twin environment. We consider a situation as a traffic conflict if the TTC of the ego vehicle falls below 3 $sec$ or if the DRAC exceeds 3 \(m/sec^2\). By simulating the ego vehicle in the traffic digital twin and continuously monitoring these Surrogate Safety Measures, we are able to identify the most hazardous locations and scenarios on a given road. This approach can provide valuable insights to transportation agencies helping them identify and address potential road safety issues. In addition, this approach can also be utilized by original equipment manufacturers (OEMs) to test advanced driver-assistance systems (ADAS) in a controlled and safe digital environment before field tests.

\section*{Results}

\subsection*{Traffic conflict identified in traffic digital twin}
\begin{figure}
    \centering
    \includegraphics[width=1\linewidth]{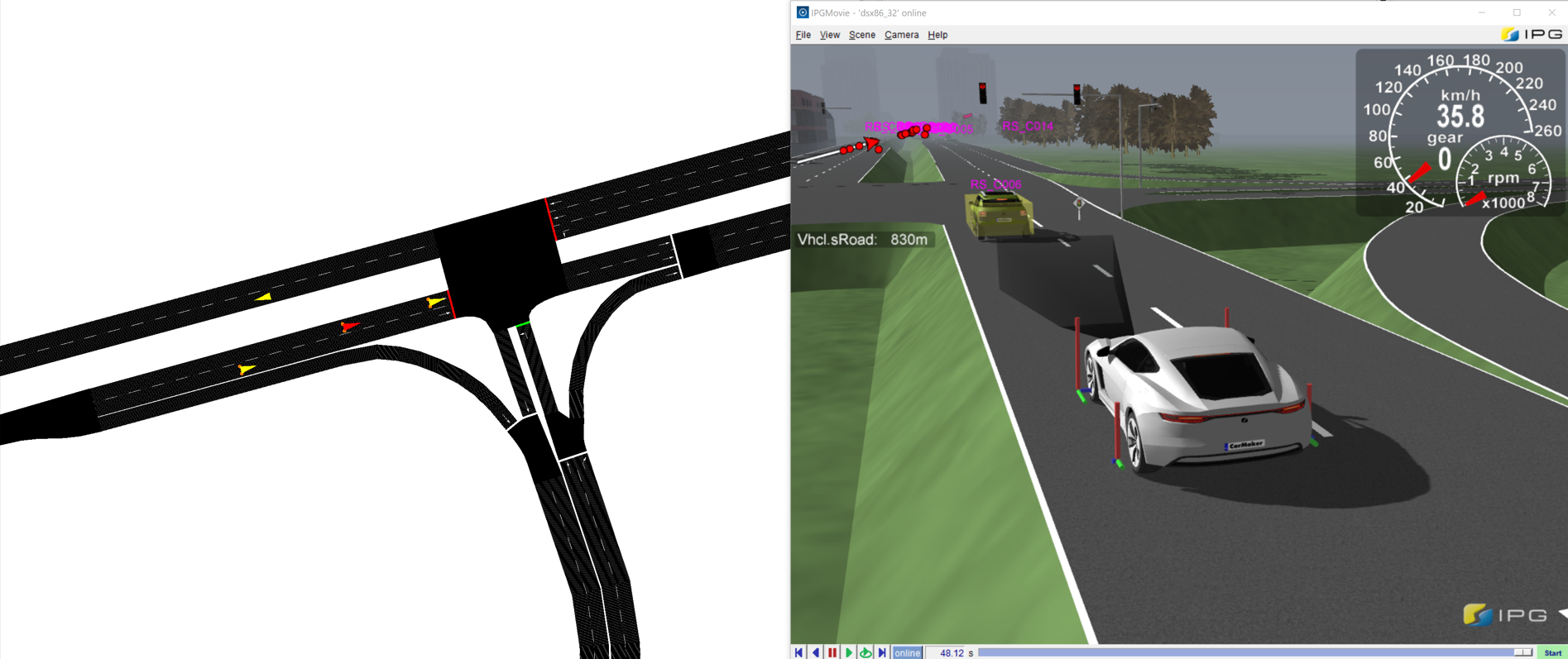}
    \caption{Co-simulation between SUMO (left) and IPG CarMaker (right)}
    \label{cosim}
\end{figure}

The generated traffic digital twin is shown in Figure \ref{cosim}, where the left-hand side shows the microscopic simulation in SUMO while the right-hand side shows the vehicle simulation in IPG CarMaker. By transmitting traffic information, signal information, vehicle dynamics, and environment information between the two software at every simulation time step, a traffic digital twin is generated. 

The baseline scenario ($\mu=0.7, L_v=150 m$) is first simulated in the generated digital twin. During a simulation, we observe a condition when ego vehicle's leader has to perform a hard brake in front of a intersection as the signal light suddenly turn red and thus the ego vehicle also has to perform a hard brake. This condition reduce TTC between the ego vehicle and its leader below 3 $sec$ and increases DRAC of the ego vehicle above 3 \(m/sec^2\), creating a traffic conflict. These traffic conflict will be looked into in detail in the following sections.

\subsection*{Impact of vehicle dynamic}
\begin{figure}[ht]
    \centering
    \begin{subfigure}[b]{0.3\textwidth}
        \centering
        \includegraphics[width=\textwidth]{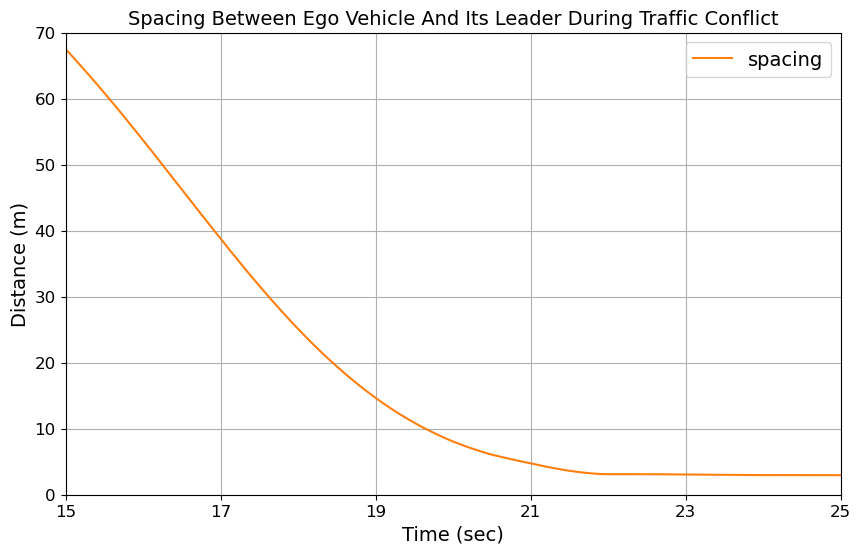}
        \caption{}
        \label{0.7x}
    \end{subfigure}
    \hfill
    \begin{subfigure}[b]{0.3\textwidth}
        \centering
        \includegraphics[width=\textwidth]{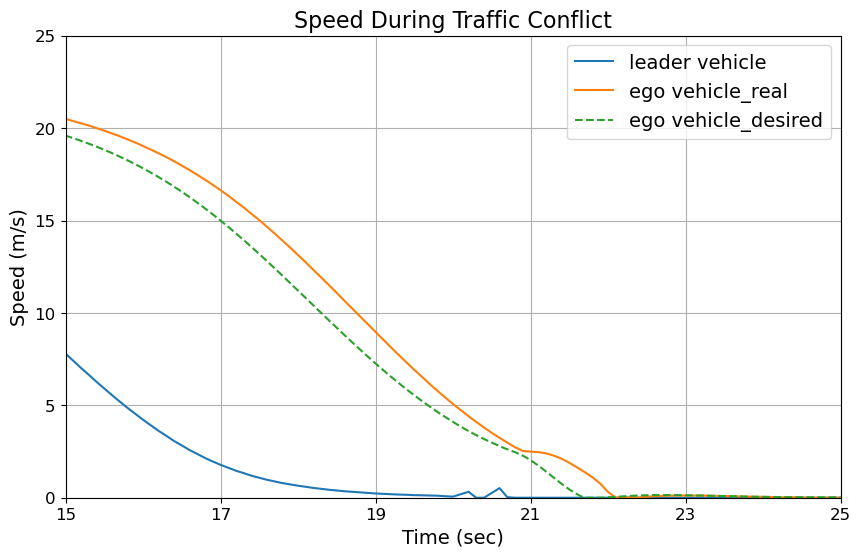}
        \caption{}
        \label{0.7v}
    \end{subfigure}
    \hfill
    \begin{subfigure}[b]{0.3\textwidth}
        \centering
        \includegraphics[width=\textwidth]{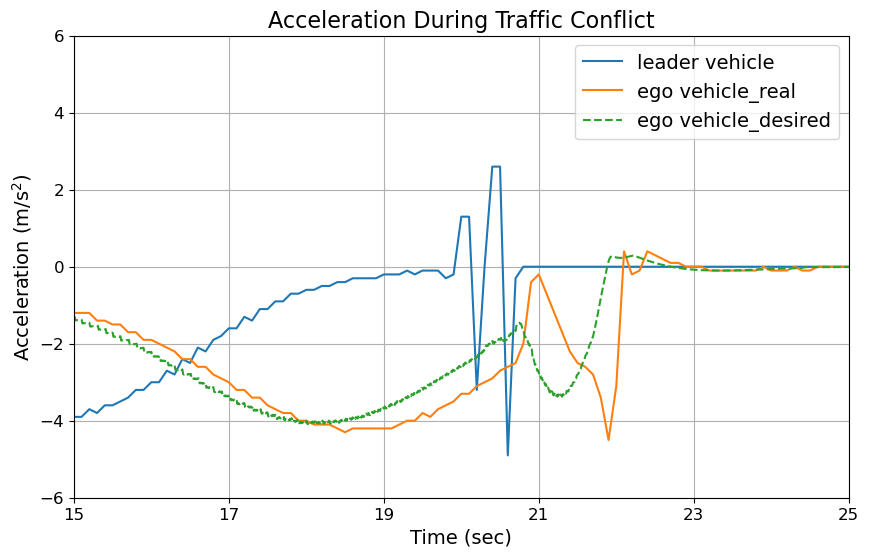}
        \caption{}
        \label{0.7a}
    \end{subfigure}
    \caption{Vehicle measurements during an identified traffic conflict: (A) spacing between ego vehicle and its leader, (B) speed of ego vehicle and its leader, and (C) acceleration of ego vehicle and its leader.}
    \label{0.7}
\end{figure}

The spacing between the ego vehicle and its leader is illustrated in Figure \ref{0.7x}. Throughout the traffic conflict, the inter-vehicle distance gradually decreased, eventually stopping with the ego vehicle about 3 meters behind its leader. The actual speeds of both the ego vehicle and the leader are shown in Figure \ref{0.7v}. Notably, the ego vehicle's \emph{desired speed} according to the IDM is also plotted as a benchmark. When the ego vehicle's low-level dynamics are incorporated into the simulation loop, its \emph{actual} speed does not match its \emph{desired speed}, and a delay between the two values is evident. As explained in \cite{wang2023cooperative}, the detailed IPG CarMaker vehicle model, which includes engine, transmission, chassis, tire forces, etc., inherently maintains a system delay. Additionally, the embedded driver model, which adjusts the brake and gas pedals based on the difference between actual and reference speeds, cannot completely eliminate the speed tracking error. The response delay introduced by the IPG CarMaker vehicle simulation module is further illustrated in Figure \ref{0.7a}. It is important to note that microscopic traffic simulators like SUMO do not account for detailed vehicle dynamics, and thus may produce overly optimistic results in collision analysis by ignoring system delays.

\subsection*{Impact of road friction}

Recognizing the significant impact of vehicle dynamic on vehicle acceleration and speed, we then investigate how this impact combined with environment factors affect traffic safety. First, we examine the impact of road friction on traffic safety. To do this, we simulate the digital twin 
under different friction coefficients ($\mu$) and identify the same traffic conflict involving the ego vehicle and its leader at the same locations. Figure \ref{friction} shows the TTC and DRAC over time during the same traffic conflict as in Figure \ref{0.7} under different friction coefficients. To clearly see when the traffic conflict starts, we start to record TTC and DRAC only when TTC is smaller than 3 $sec$ or DRAC is larger than 3 $m/sec^2$. In addtion, since TTC tends to infinity as the speed difference between the ego and leader vehicle reduces to 0, we only plot TTC values that are between 0 to 5 $sec$. As shown in Figure \ref{friction}, in all three cases, the traffic conflict starts at around 16.5 $sec$. At around 21-22 $sec$, TTCs starts to dramatically increase and DRAC stay 0 indicating that the ego vehicle successfully stops to avoid an rear-end crash with its leader. It is evident that as $\mu$ decreases, the minimum TTC reduces and maximum DRAC increases signaling a higher crash risk.

\begin{figure}[ht]
    \centering
    \begin{subfigure}[b]{0.32\textwidth}
        \centering
        \includegraphics[width=\textwidth]{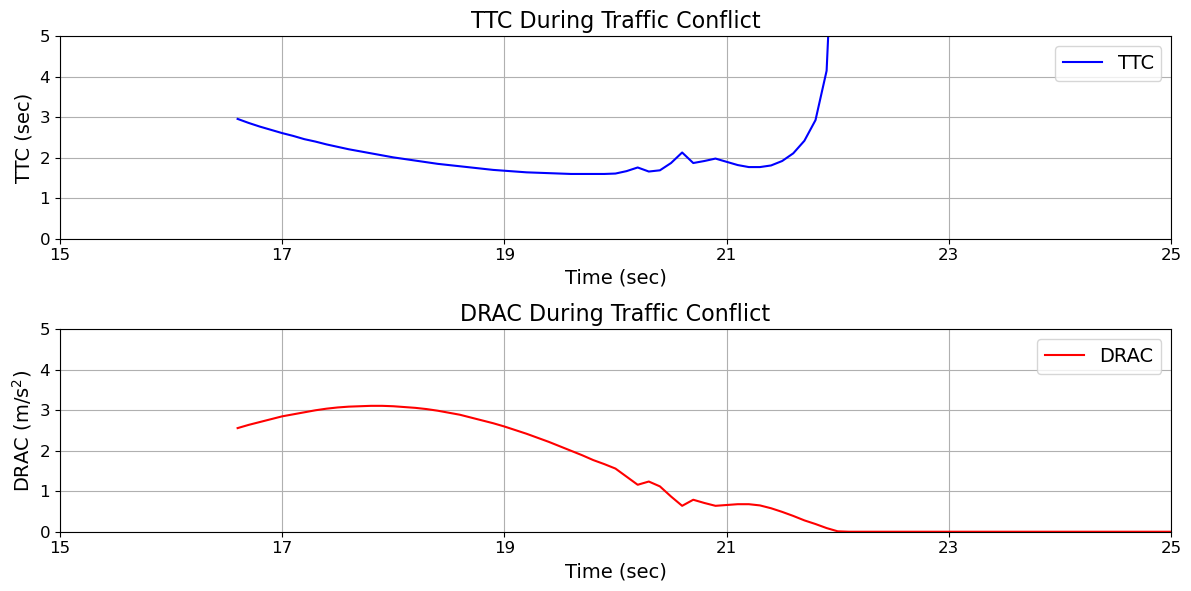}
        \caption{}
        \label{0.4x}
    \end{subfigure}
    \hfill
    \begin{subfigure}[b]{0.32\textwidth}
        \centering
        \includegraphics[width=\textwidth]{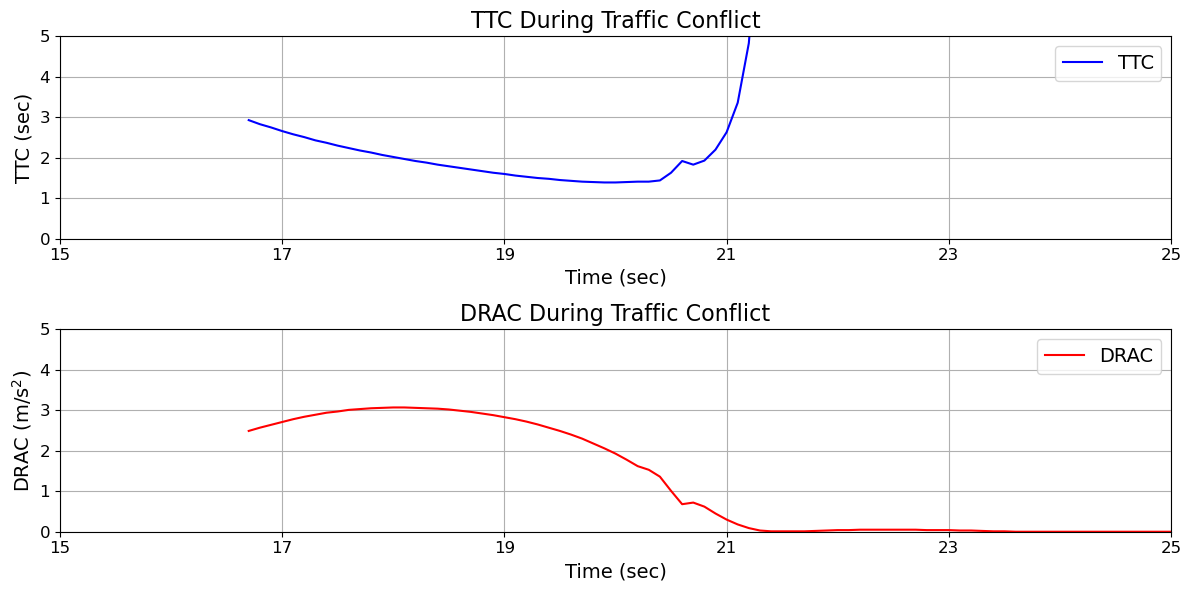}
        \caption{}
        \label{0.4v}
    \end{subfigure}
    \hfill
    \begin{subfigure}[b]{0.32\textwidth}
        \centering
        \includegraphics[width=\textwidth]{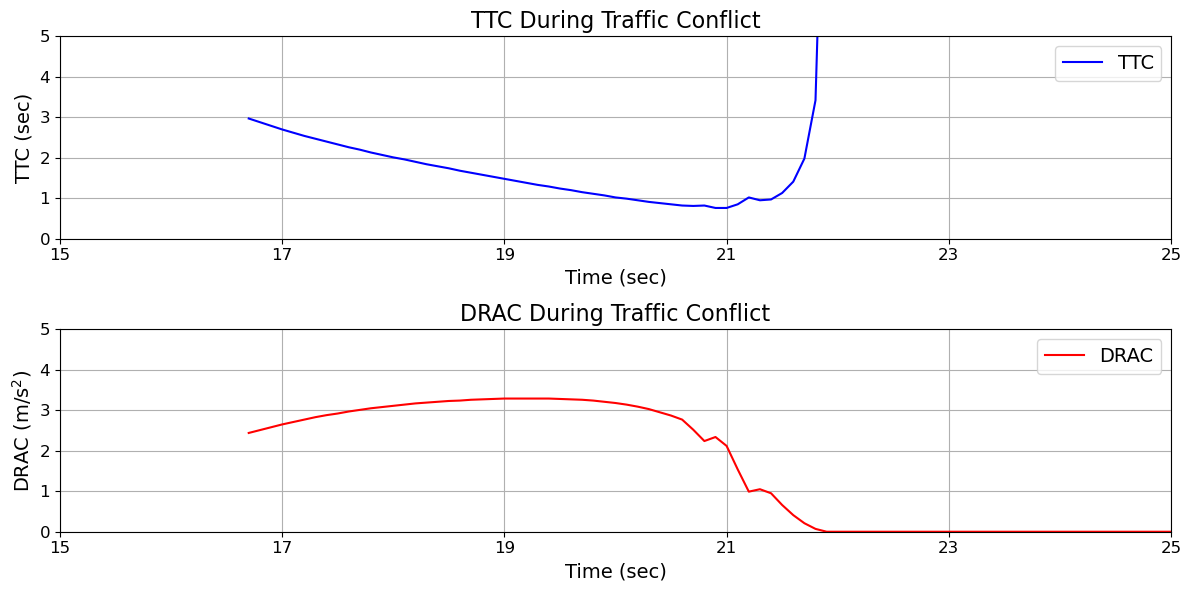}
        \caption{}
        \label{0.4t}
    \end{subfigure}
    \caption{Time to Collision (TTC) and Deceleration Rate to Avoid a Crash (DRAC) during the same traffic conflict under different friction coefficients($\mu$): (A) $\mu=0.7$, (B) $\mu=0.4$, and (C) $\mu=0.35$.}
    \label{friction}
\end{figure}

Table \ref{friction} summarizes the mean and standard deviation of the minimum TTC and the maximum DRAC during the same traffic conflict across 10 simulation runs under different friction coefficients ($\mu$). The friction coefficients analyzed are $\mu = 0.7$, $\mu = 0.4$, and $\mu = 0.35$. It can be observed that as ($\mu$) decreases from 0.7 to 0.35, mean of minimum TTC significantly reduces from 1.6040 sec to 0.7010 sec which indicates that vehicles have less time to avoid a collision in lower friction conditions. Concurrently, standard deviation of minimum TTC increases from 0.0092 sec to 0.0474 sec suggesting greater variability in collision avoidance times under more slippery conditions. This increased variability indicates increased unpredictability when the road is wet, which raise safety concerns. On the other hand, mean of maximum DRAC increases as the ($\mu$) decreases. This trend indicates that higher deceleration rates are required to avoid crashes when the road is more slippery. Similar to standard deviation of minimum TTC, standard deviation of maximum DRAC also rises as ($\mu$) decreases, reflecting more variability in the deceleration rates needed under lower friction conditions.

These results highlight the extended capability of digital twin in analyzing traffic safety compared to traditional microscopic traffic simulation. By simulating different friction conditions in a digital twin which correspond to different weather conditions, we can identify critical scenarios and locations that has safety concerns and thereby improve traffic controls strategies and transportation infrastructures to mitigate the risks of crashes. For instance, in the results shown above, we identify a potential crash at a signalized intersection when road friction decreases due to weather conditions such as rain or snow. This information can be used to provide real-time feedback to traffic signal controllers, recommending a longer green time to let the driver traverse the intersection or a longer yellow time to allow drivers more time to react and stop safely at the intersection.

\begin{table}[ht]
\centering
\begin{tabular}{|l|c|c|c|}
\hline
 & $\mu = 0.7$ & $\mu = 0.4$ & $\mu = 0.35$ \\
\hline
mean of min TTC ($sec$) & 1.6040 & 1.3670 & 0.7010 \\
\hline
standard deviation of min TTC ($sec$) & 0.0092 & 0.0101 & 0.0474 \\
\hline
mean of max DRAC ($m/sec^2$) & 3.0940 & 3.1500 & 3.3300 \\
\hline
standard deviation of max DRAC ($m/sec^2$) & 0.0092 & 0.0100 & 0.0272 \\
\hline
\end{tabular}
\caption{\label{friction}Surrogate Safety Measures (SSMs) under scenarios with different friction coefficients($\mu$). Each scenario is simulated 10 times with default visibility ($Lv=150m$). The mean and standard deviation of minimum Time to Collision (TTC) and maximum Deceleration Rate to Avoid a Crash (DRAC) across 10 simulations are then calculated.}
\end{table}

\subsection*{Impact of visibility}

Next, we look into the impact of visibility on traffic safety. Figure \ref{visibility} shows the TTC and DRAC over time during the same traffic conflict under different visibility values, while Table \ref{visibility} summarize mean and standard deviation of minimum TTC and maximum DRAC from 10 simulation runs. In our simulated digital twin, visibility($L_v$) does not have a great impact on the SSMs when $L_v$ is larger than 55 m. However, as visibility decreases from 55 m to 47 m, both TTC and DRAC changes much faster over time during the traffic conflict shown in Figure \ref{visibility}, indicating that a harder brake is needed due to less distance to react. Meanwhile, mean of minimum TTC from 10 simulation runs significantly reduces from 1.6050 $sec$ to 0.8140 $sec$ and mean of maximum DRAC moves from 3.0901 $m/sec^2$ at $L_v = 55m$ to 3.8400 $m/sec^2$ at $L_v = 47m$, as shown in Table \ref{visibility}. This reduction also indicates that drivers have less time to react and avoid a crash under lower visibility conditions. Correspondingly, standard deviation of minimum TTC and maximum DRAC increases, suggesting greater variability and unpredictability in crash avoidance times when visibility is poor. In addition, multiple scenarios are also simulated with $L_v$ lower than 47 m. Results show that a rear-end crash between the ego vehicle and its leader will always occur at the same location. This indicates that a safety treatment (e.g., lower speed limit or more lighting during night) is needed at this location to increase the traffic safety on this road. These findings emphasize the impact of visibility on traffic safety and illustrate the capability of  traffic digital twins in simulating and evaluating these impacts which traditional methods cannot.

\begin{figure}[ht]
    \centering
    \begin{subfigure}[b]{0.32\textwidth}
        \centering
        \includegraphics[width=\textwidth]{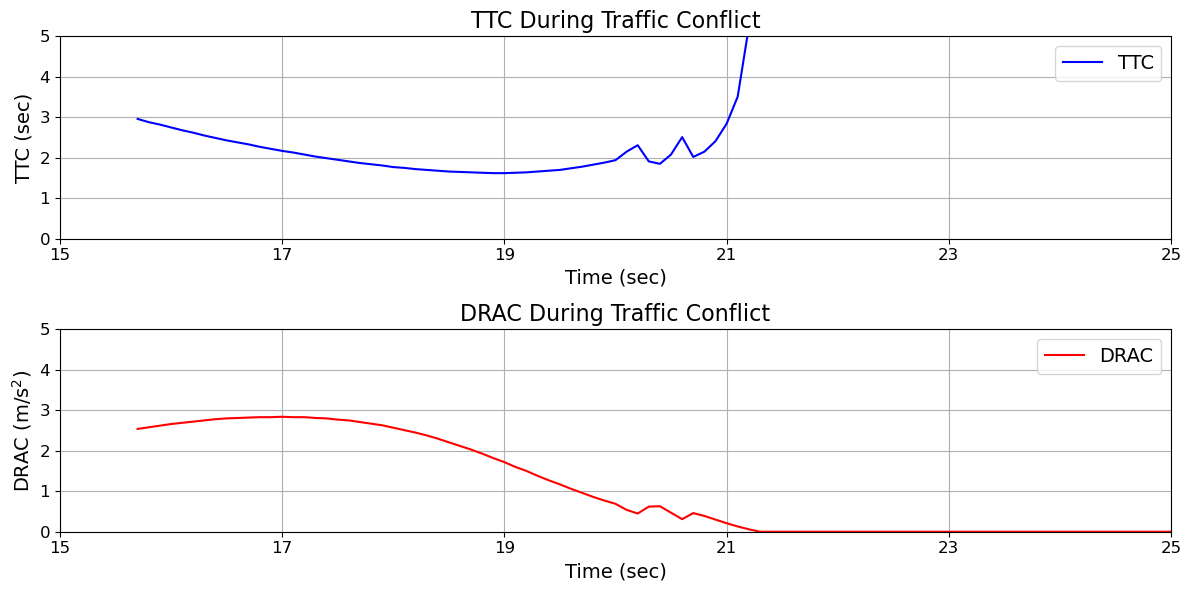}
        \caption{}
        \label{0.4x}
    \end{subfigure}
    \hfill
    \begin{subfigure}[b]{0.32\textwidth}
        \centering
        \includegraphics[width=\textwidth]{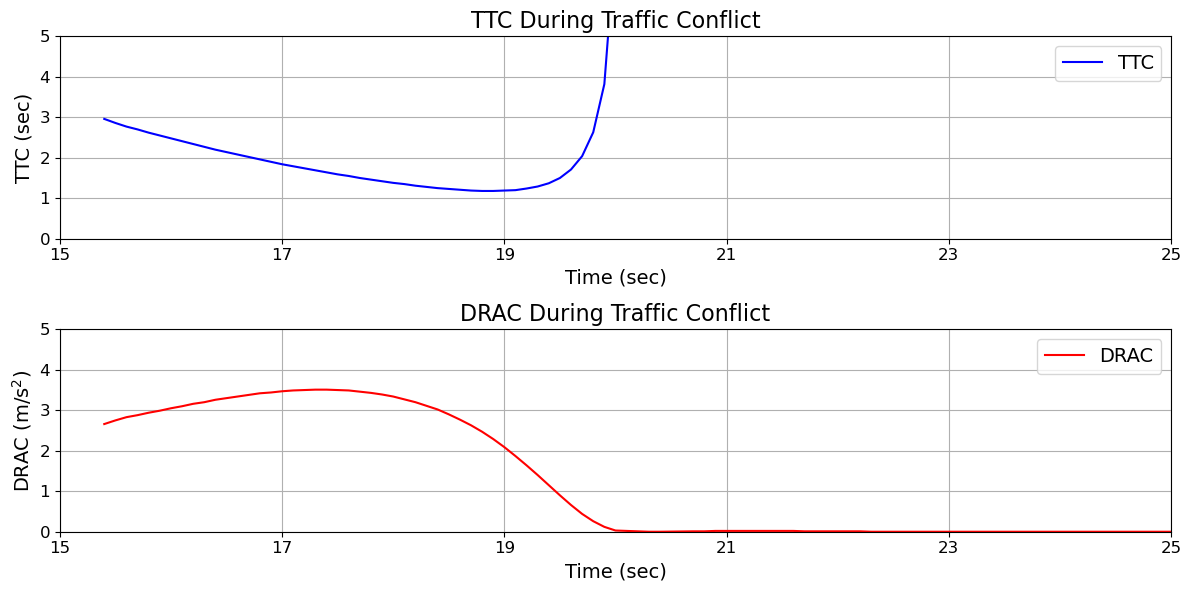}
        \caption{}
        \label{0.4v}
    \end{subfigure}
    \hfill
    \begin{subfigure}[b]{0.32\textwidth}
        \centering
        \includegraphics[width=\textwidth]{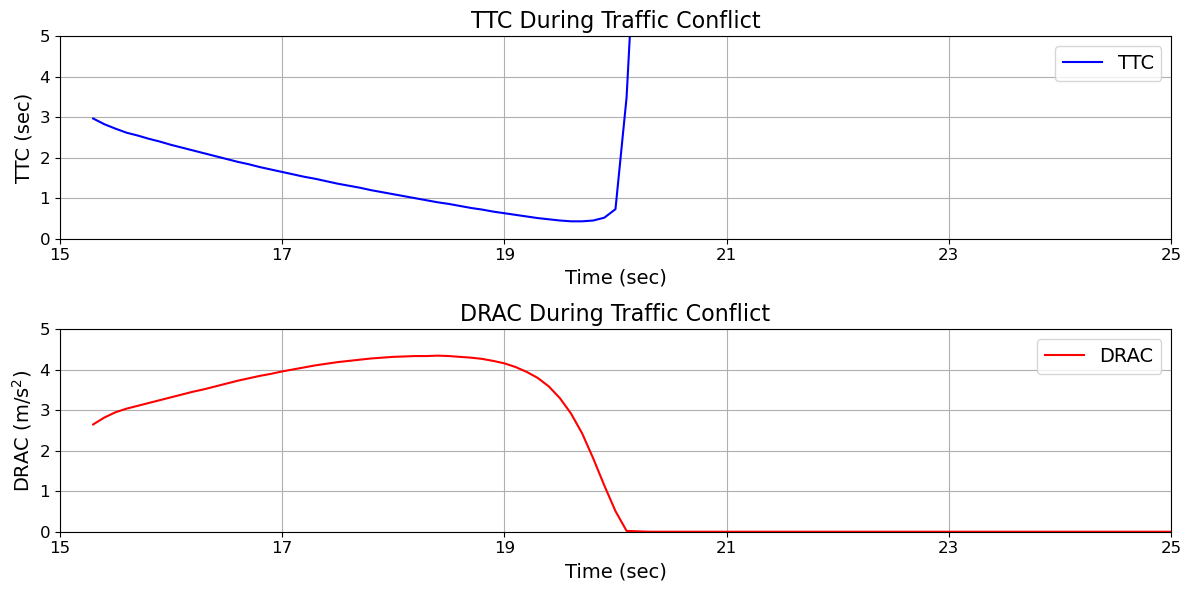}
        \caption{}
        \label{0.4t}
    \end{subfigure}
    \caption{Time to Collision (TTC) and Deceleration Rate to Avoid a Crash (DRAC) during the same traffic conflict under different visibility ($L_v$): (A) $L_v=55 m$, (B) $L_v=50 m$, and (C) $L_v=47 m$.}
    \label{0.4}
\end{figure}

\begin{table}[ht]
\centering
\begin{tabular}{|l|c|c|c|}
\hline
 & $L_v = 55m$ & $L_v = 50m$ & $L_v = 47m$ \\
\hline
mean of min TTC ($sec$) & 1.6050 & 1.3940 & 0.8140 \\
\hline
standard deviation of min TTC ($sec$) & 0.0082 & 0.1139 & 0.2001 \\
\hline
mean of max DRAC ($m/sec^2$) & 3.0901 & 3.2120 & 3.8400 \\
\hline
standard deviation of max DRAC ($m/sec^2$) & 0.0432 & 0.1589 & 0.2659 \\
\hline
\end{tabular}
\caption{\label{visibility} Surrogate Safety Measures (SSMs) under scenarios with different visibility($L_v$). Each scenario is simulated 10 times with a fixed friction coefficient of 0.7. The mean and standard deviation of minimum Time to Collision (TTC) and maximum Deceleration Rate to Avoid a Crash (DRAC) across 10 simulations are then calculated.}
\end{table}

\section*{Discussion}

This study demonstrates the significant potential of digital twin in enhancing traffic safety analysis. A digital twin is created for a road segment in Tuscaloosa, Alabama, providing a realistic platform for simulating various traffic scenarios. The simulated scenarios incorporate different weather conditions and visibility levels using Surrogate Safety Measures (SSMs) such as Time to Collision (TTC) and Deceleration Rate to Avoid a Crash (DRAC) to assess traffic safety  under these varying conditions. Results indicate that traffic digital twin can identify critical safety concerns that conventional microscopic traffic simulation cannot. 

In summary, the integration of digital twin technology into traffic safety analysis represents a forward step in our ability to predict and reduce traffic safety issues. By leveraging the capabilities of digital twins, transportation agencies can enhance their understanding of traffic systems, identify locations with potential safety concerns, and improve traffic control strategies and transportation infrastructure to enhance traffic safety.

\section*{Data availability}
The datasets used  during the current study are available from the corresponding author on reasonable request.

\bibliography{reference}

\section*{Acknowledgements}

This material is based upon work supported by the US Department of Energy, Vehicle Technologies Office, Energy Efficient Mobility Systems (EEMS) program, under project \textit{Improving Network-Wide Fuel Economy and Enabling Traffic Signal Optimization Using Infrastructure and Vehicle-Based Sensing and Connectivity} (EEMS107).

\section*{Author contributions statement}

Conceptualization: G.X., J.C., Z.W., A.Z., M.S., J.B., Y.S.; Methodology: G.X., J.C., Z.W., A.Z., M.S., J.B., Y.S.; Result analysis: G.X., J.C., Z.W.; Writing and editing: G.X., J.C., Z.W., M.S., Y.S.

\section*{Competing interests}

The authors declare no competing interests.

\end{document}